\documentclass[twocolumn,showpacs,preprintnumbers,amsmath,amssymb]{revtex4}

\newcommand{\nc}{\newcommand}
\nc{\rnc}{\renewcommand}
\nc{\beq}{\begin{equation}}
\nc{\eeq}{\end{equation}}
\nc{\nn}{\nonumber}
\rnc{\(}{\left(}
\rnc{\)}{\right)}
\rnc{\[}{\left[}
\rnc{\]}{\right]}
\rnc{\i}{{\rm i}}
\nc{\db}{\displaybreak[0]\\}

\nc{\bra}{\langle}
\nc{\ket}{\rangle}

\usepackage{graphicx}
\usepackage{dcolumn}
\usepackage{bm}

\begin{document}

\preprint{}

\title{Density matrix for the \\ 
kink ground  state of the ferromagnetic XXZ chain}

\author{Kohei Motegi}
 \email{motegi@gokutan.c.u-tokyo.ac.jp}
\author{Kazumitsu Sakai}%
 \email{sakai@gokutan.c.u-tokyo.ac.jp}
\affiliation{%
Institute of physics, University of Tokyo, \\
Komaba 3-8-1, Meguro-ku, Tokyo 153-8902, Japan
}%

\date{December 30, 2008}

\begin{abstract}
The exact expression for the density matrix 
of the kink  ground state of the 
ferromagnetic XXZ chain is obtained.
Utilizing this, we exactly calculate various 
correlation functions such as the longitudinal and
transverse spin-spin correlation functions, and  
the ferromagnetic and antiferromagnetic string 
formation probabilities.
The asymptotic behaviors of these correlation functions
are  also analyzed. As a consequence, we find that
the spin-spin correlation functions decay exponentially 
for large distances, while the string 
formation probabilities exhibit Gaussian decay
for large strings. 
We also evaluate the entanglement entropy,
which shows interesting behaviors due to the lack of 
the translational invariance of the state. 
\end{abstract}

\pacs{05.30-d, 75.10.Pq, 02.30.Ik}


\maketitle
%
\section{Introduction}
%
The effects of quantum fluctuations of interacting 
quantum systems can be investigated by studying the
correlation functions or quantifying the entanglement 
of the system, which are currently paid much attention.
The exact evaluation of the correlation functions,
however, is still a challenging problem even when models
are completely integrable.
The spin-1/2 XXZ model in one-dimension is one of the 
most fundamental model, which can be  exactly 
solved by the Bethe ansatz.
As concerns for the correlation functions in the {\it 
antiferromagnetic} ground state, a few exact results 
are known so far: several short distance spin-spin correlation 
functions (see \cite{SS} and references therein) and the ferromagnetic 
string formation probability 
(which is the probability to find a ferromagnetic string with
certain length) for $\Delta=1/2$ ($\Delta$: anisotropy parameter)
\cite{RS,KMST}.

In this paper, we intensively consider the correlation 
functions for the {\it ferromagnetic} regime of the XXZ chain.
In this regime, it has been well-known that there are 
two translationally invariant ground states \textit{up}
and \textit{down}, the state with all spins up
and the state with all spins down. In addition to these 
trivial ground states, two classes of non-translationally
invariant ground states \textit{kink} and \textit{antikink}
were found in \cite{GW} (see also \cite{ASW} for
{\it finite} XXZ chain with boundary magnetic field).
Though it is not obvious that the \textit{kink}
is the ground state in the infinite lattice limit,
the authors in \cite{GW} proved it under the assumption
that the ground states should be ``frustration free",
i.e. minimize not only the energy of the total 
Hamiltonian, but also the energy of the local 
Hamiltonian.
Furthermore, it was shown that the ``frustration free" 
ground states \textit{up}, \textit{down}, \textit{kink} and 
\textit{antikink} are the complete set of the ground states
\cite{M, KN}. In \cite{KN2}, the exact value of 
the spectral gap was obtained and shown to be independent 
of the reference ground state.

More recently, from the interest in the correspondence
between the ground state of the ferromagnetic XXZ chain 
and the
``quantum" Hamiltonian of the crystal melting model,
some special correlation functions such as
the magnetization and the longitudinal spin-spin
correlation function of the kink ground state were 
exactly calculated \cite{DOR}.

Utilizing the generating function developed in \cite{GW}, 
in this paper, we derive the exact expression of the density 
matrix for the kink ground state of the ferromagnetic 
XXZ chain.
By using this, various correlation functions can
be systematically calculated for arbitrary interaction
strengths and for arbitrary distances. The following
correlation functions are particularly calculated
here: the transverse and the longitudinal
spin-spin correlation functions,
the ferromagnetic (antiferromagnetic) string
formation probabilities which are the probability
finding a ferromagnetic (antiferromagnetic) string
in the kink ground state. The entanglement entropy
of the system is also evaluated (see \cite{ASS}
for finite XXZ chain with boundary magnetic field).
Analyzing the asymptotic behaviors of these
correlation functions, we find that the the
spin-spin correlation functions exponentially
decay for the large distances. On the other
hand, both the ferromagnetic and the antiferromagnetic
string formation probabilities exhibit Gaussian 
decay for large strings. To authors' knowledge,
this study is the first to investigate systematically
the correlation functions and their asymptotics
in the kink ground state.

This paper is organized as follows. In the next section, 
the kink ground state of the infinite XXZ chain in 
the ferromagnetic regime is considered. By using the 
generating function, the exact expression for the density 
matrix is derived.
From this, we concretely analyze various correlation 
functions in section \text{III}. The asymptotic behavior
of the correlation functions are discussed in section 
\text{IV}. Section $\textrm{V}$ is devoted to conclusion.
%
\section{density matrix  of the kink ground state}
%
In this section, we derive the exact expression for the 
density matrix  of the infinite ferromagnetic XXZ chain
in the kink ground state.
The Hamiltonian is defined by
\begin{equation}
H=- \sum_{m \in \mathbb{Z}}
\left[
\sigma_{m}^x \sigma_{m+1}^x+\sigma_{m}^y \sigma_{m+1}^y
+\Delta
(\sigma_{m}^z \sigma_{m+1}^z-1)
\right],
\end{equation}
where $\sigma_m^{\alpha}$, $\alpha=x,y,z$ are the Pauli 
matrices acting on the \textit{m}th site and 
$\Delta$ is the anisotropy parameter. We shall consider 
the ferromagnetic regime $\Delta>1$. Parametrizing $\Delta$ as
\begin{equation}
\Delta=\frac{q^{\frac{1}{2}}+q^{-\frac{1}{2}}}{2},
\end{equation}
$\Delta>1$ corresponds to $0<q<1$.

It is known that there are infinitely many 
zero energy kink ground states
interpolating between spin up at $-\infty$
and spin down at $\infty$ \cite{GW,ASW}.
A kink ground state is the superposition
of kinks which have the same center.
The center of the kink is defined as the 
half integer-valued position where
the number of up spins on the right of it
is equal to the number of down spins on the left
of it.
Denote the kink ground state whose center is 
at $j-\frac{1}{2}$ $(j \in \mathbb{Z})$ by $| \Psi_j \ket$.
Any $| \Psi_j \ket$ can be extracted from the following 
generating function \cite{GW},
\begin{align}
| \Psi(z) \ket=&
  \bigotimes_{x \in \mathbb{Z}_{<0}} 
   (| \uparrow \ket_{x}+z^{-1} q^{-\frac{1}{2}(\frac{1}{2}+x)}
    | \downarrow  \ket_{x})                         \nn \\
&\otimes
  \bigotimes_{y \in \mathbb{Z}_{\geq 0}} 
   (| \downarrow \ket_{y}+z q^{\frac{1}{2}(\frac{1}{2}+y)}
    | \uparrow  \ket_{y}).
\label{GF}
\end{align}
$|\Psi_j \ket$ is the coefficient of $z^{j}$ of the expansion of
$| \Psi(z) \ket$, i.e,
\begin{equation}
| \Psi(z) \ket=\sum_{j \in \mathbb{Z}} z^j |\Psi_j \ket.
\end{equation}

Let us calculate the form factors,
\begin{equation}
_{k} \bra \prod_{j=1}^n E_{x_j}^{\epsilon_j^{\prime} \epsilon_j}  \ket_{l}
:=\frac{
          \bra \Psi_k | \prod_{j=1}^n E_{x_j}^{\epsilon_j^{\prime} \epsilon_j} | \Psi_l \ket
       }
       {
            \bra \Psi_k | \Psi_k \ket^{\frac{1}{2}}
            \bra \Psi_l | \Psi_l \ket^{\frac{1}{2}}
},
\label{exp0}
\end{equation}
where $\{\epsilon_j\}$, $\{\epsilon_j^{\prime}\}\in\{+,-\}$ and
$E_j^{\pm \pm}=(1 \pm \sigma_j^z)/2,
E_j^{\pm \mp}=\sigma_j^{\pm}=(\sigma_j^x \pm \i \sigma_j^y)/2$.
$x_j$ is the position of the site where the operator 
$E_{x_j}^{\epsilon_j^{\prime} \epsilon_j}$ acts on, and 
is assumed to be $x_j \neq  x_k$ for $j \neq k$.
First we calculate the norm $\bra \Psi_n | \Psi_n \ket$
appearing in the denominator of \eqref{exp0}.
It can be obtained by calculating $\bra \Psi(z) | \Psi(z) \ket$:
\begin{align}
\bra \Psi(z) | \Psi(z) \ket=&(-wq^{\frac{1}{2}};q)_{\infty}
(-w^{-1}q^{\frac{1}{2}};q)_{\infty} \nn \\
=& \frac{1}{(q;q)_{\infty}} \sum_{j=-\infty}^{\infty}w^j q^{\frac{j^2}{2}},
\label{comp1}
\end{align}
where $w=z^2$ and $(a;q)_{\infty}:=\prod_{j=0}^{\infty}(1-a q^{j})$. 
In the second equality, we have used the Jacobi triple
product identity,
\begin{eqnarray}
(q;q)_{\infty} (-wq^{\frac{1}{2}};q)_{\infty}
(-w^{-1}q^{\frac{1}{2}};q)_{\infty}
=\sum_{j=-\infty}^{\infty}w^j q^{\frac{j^2}{2}}.
\end{eqnarray}
Noting 
$
\bra \Psi(z) | \Psi(z) \ket= \sum_{j=-\infty}^{\infty} w^j \bra \Psi_j | \Psi_j \ket, 
$
we have
\begin{equation}
\bra \Psi_j | \Psi_j \ket=\frac{q^{\frac{j^2}{2}}}{(q;q)_{\infty}}.
\label{norm}
\end{equation}
Next we compute 
$ \bra \Psi_k | \prod_{j=1}^n E_{x_j}^{\epsilon_j^{\prime} \epsilon_j} | \Psi_l \ket $ 
appearing in the numerator of \eqref{exp0}. Note that
$ \bra \Psi_k | \prod_{j=1}^n E_{x_j}^{\epsilon_j^{\prime} \epsilon_j} | \Psi_l \ket = 0$
unless $l-k=\delta$
where 
\begin{equation}
\delta:=\sharp\{j;(\epsilon_j,\epsilon'_j)=(+,-) \}-
       \sharp\{j;(\epsilon_j,\epsilon'_j)=(-,+) \}.
\end{equation}
Then one can see
\begin{align}
&\bra \Psi(z) |
\prod_{j=1}^n E_{x_j}^{\epsilon_j^{\prime} \epsilon_j}
| \Psi(z) \ket=\sum_i w^\frac{i}{2} \bra \Psi_{\frac{i-\delta}{2}} |
\prod_{j=1}^n E_{x_j}^{\epsilon_j^{\prime} \epsilon_j} |
\Psi_{\frac{i+\delta}{2}}  
\ket, \nn \\
&| \Psi_{j \not\in \mathbb{Z}} \ket:=0. \label{exp1} 
\end{align}
By induction, one can show that the following holds:
\begin{align}
\bra & \Psi(z) | \prod_{j=1}^n E_{x_j}^{\epsilon_j^{\prime} \epsilon_j} | \Psi(z) \ket
=\frac{1}{(q;q)_{\infty}} 
\sum_{j=-\infty}^{\infty}w^j q^{\frac{j^2}{2}} \nn \\
&\times \prod_{j=1}^n (w \zeta_j)^{\[ \epsilon_j^{\prime} \epsilon_j \]}
\sum_{k=0}^{\infty}(-w)^k 
\sum_{j=1}^n \frac{\zeta_j^{k+n-1}}{\prod_{l \neq j}^n(\zeta_j-\zeta_l)},
 \label{exp2}
\end{align}
where $\zeta_j=q^{\frac{1}{2}+x_j}$ and 
$\[ ++ \]=1, \[--\]=0, \[+- \]=\[-+ \]=\frac{1}{2}$
(see Appendix for the proof of \eqref{exp2}).
Comparing \eqref{exp1} and \eqref{exp2}, we obtain
\begin{align}
\bra \Psi_{\frac{i-\delta}{2}} |
&\prod_{j=1}^n E_{x_j}^{\epsilon_j^{\prime} \epsilon_j} |
\Psi_{\frac{i+\delta}{2}} 
\ket =\frac{1}{(q;q)_{\infty}}
\prod_{j=1}^n (\zeta_j)^{\[ \epsilon_j^{\prime} \epsilon_j \]}
\sum_{k=0}^{\infty}(-1)^k
\nn \\
&\times
\sum_{j=1}^n \frac{\zeta_j^{k+n-1}}{\prod_{l \neq j}^n(\zeta_j-\zeta_l)}
q^{\frac{(\frac{i}{2}-k-
\sum_{m=1}^n \[ \epsilon_m^{\prime} \epsilon_m \])^2}{2}}.
\label{exp3}
\end{align}
Combining \eqref{exp0}, \eqref{norm} and \eqref{exp3}, 
we finally arrive at
\begin{align}
_{\frac{i-\delta}{2}} &\bra \prod_{j=1}^n 
E_{x_j}^{\epsilon_j^{\prime} \epsilon_j}  
\ket_{\frac{i+\delta}{2}}=
\prod_{j=1}^n (\zeta_j)^{\[ \epsilon_j^{\prime} \epsilon_j \]}
\sum_{k=0}^{\infty}(-1)^k \nn \\
&\times \sum_{j=1}^n \frac{\zeta_j^{k+n-1}}{\prod_{l \neq j}(\zeta_j-\zeta_l)}
q^{\frac{(\frac{i}{2}-k-\sum_{m=1}^n 
\[ \epsilon_m^{\prime} \epsilon_m
 \])^2}{2}-\frac{i^2+\delta^2}{8}}.
\label{density}
\end{align}
In particular, when the operator 
$\prod_{j=1}^n E_{x_j}^{\epsilon_j^{\prime} \epsilon_j} $ 
preserves the total spin, i.e. $\delta=0$,
and the center of the kink is at $-1/2$, i.e. $i=0$,
the density matrix elements 
of the kink ground state whose center 
is located at $-1/2$ is given by
\begin{align}
\bra & \prod_{j=1}^n E_{x_j}^{\epsilon_j^{\prime}  \epsilon_j}  
\ket
:=
_{0} \bra \prod_{j=1}^n E_{x_j}^{\epsilon_j^{\prime} \epsilon_j}  
\ket_{0}
=\prod_{j=1}^n (\zeta_j)^{\[ \epsilon_j^{\prime} \epsilon_j \]}
\nn \\
&\times \sum_{k=0}^{\infty}(-1)^k 
\sum_{j=1}^n \frac{\zeta_j^{k+n-1}}{\prod_{l \neq j}^n(\zeta_j-\zeta_l)}
q^{\frac{(k+\sum_{m=1}^n \[ \epsilon_m^{\prime} \epsilon_m \])^2}{2}}.
\label{exp4}
\end{align}
%
%
\section{correlation functions}
%
Here we analyze several crucial correlation
functions: the magnetization, the longitudinal
and transverse spin-spin correlation functions,
and the ferromagnetic (antiferromagnetic) 
string formation probability which is the
probability finding a ferromagnetic 
(antiferromagnetic) string in the kink ground state.
At the end of this section, the entanglement entropy
of the system is also considered.
These correlation functions are directly calculated 
by the density matrix \eqref{exp4} derived in the 
preceding section.
Note that the correlation functions discussed here
are for the kink ground state whose center is
located at $-1/2$. Other cases can also be treated
by using \eqref{density}.
%
\begin{figure}[]
\includegraphics[width=\linewidth]{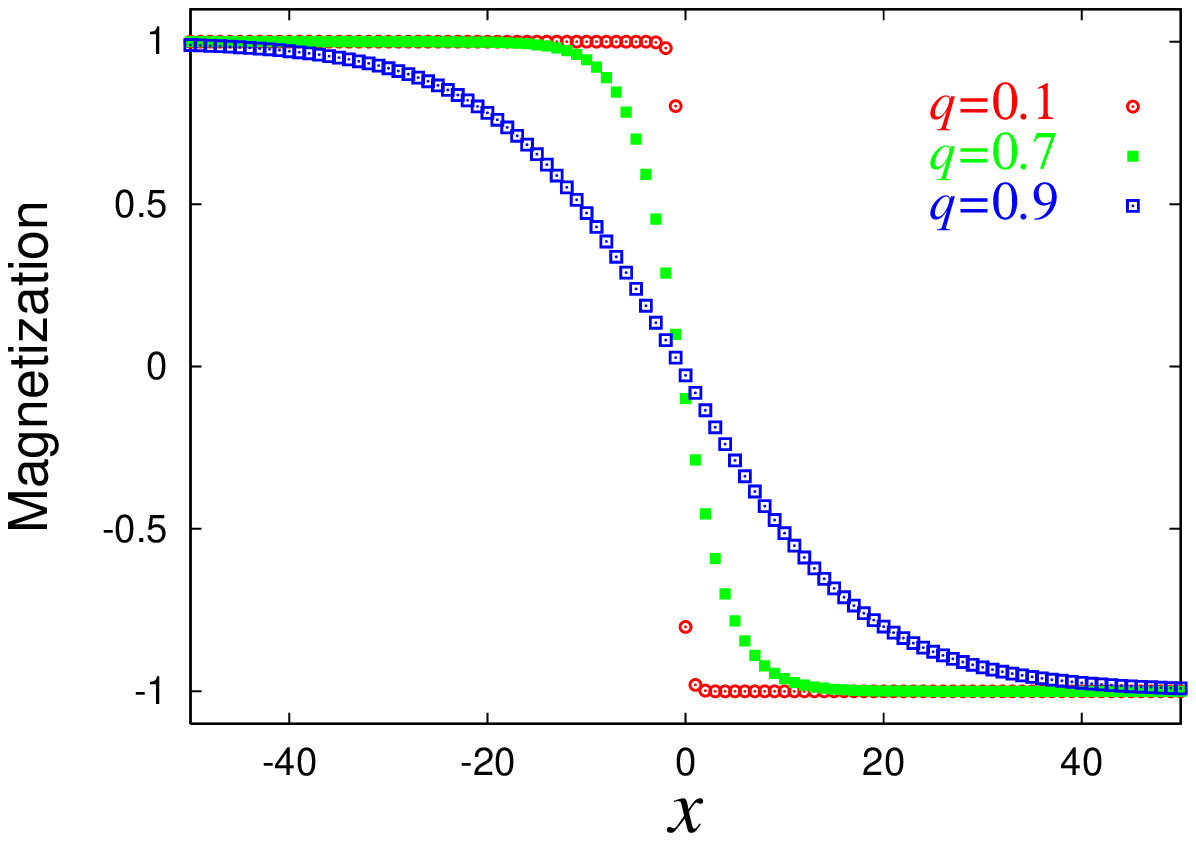}
\caption{Magnetization $\bra \sigma_x^z \ket$.}
\label{magnet}
\end{figure}
%

Let us list the explicit expressions of these correlation
functions.\\
\noindent
(i) Magnetization \cite{DOR}:
\begin{equation}
\bra \sigma_x^z \ket
=1-2 \sum_{k=0}^{\infty}(-1)^k
q^{\frac{k^2}{2}}
q^{k(x+\frac{1}{2})}.
\label{magnetization}
\end{equation}
(ii) Longitudinal spin-spin correlation function \cite{DOR}:
\begin{align}
\bra & \sigma_{x_1}^{z} \sigma_{x_2}^{z} \ket
=1+2 \sum_{k=0}^{\infty}(-1)^{k+1} 
q^{\frac{(k+1)^2}{2}} \nn \\
& \times \frac{(q^{x_1+\frac{1}{2}}+q^{x_2+\frac{1}{2}})
(q^{(k+1)(x_1+\frac{1}{2})}-q^{(k+1)(x_2+\frac{1}{2})})}
{q^{x_1+\frac{1}{2}}-q^{x_2+\frac{1}{2}}}.
\label{z-correlation}
\end{align}
(iii) Transverse spin-spin correlation function:
\begin{align}
\bra \sigma_{x_1}^{+} \sigma_{x_2}^{-} \ket
=& q^{\frac{1}{2}(x_1+\frac{1}{2})}
q^{\frac{1}{2}(x_2+\frac{1}{2})}
\sum_{k=0}^{\infty}
(-1)^k q^{\frac{(k+1)^2}{2}} \nn \\
&\times \frac{q^{(k+1)(x_1+\frac{1}{2})}-q^{(k+1)(x_2+\frac{1}{2})}}
{q^{x_1+\frac{1}{2}}-q^{x_2+\frac{1}{2}}}.
\label{pm-correlation}
\end{align}
(iv) Ferromagnetic String Formation Probability:
\begin{align}
&P_{\rm f}(x,n):=\bra E_x^{++}  
\cdots  E_{x+n-1}^{++} \ket \nn \\
&=q^{nx+\frac{n^2}{2}} \sum_{k=0}^{\infty}(-1)^k
q^{\frac{(k+n)^2}{2}} \sum_{j=1}^n
\frac{q^{(x+j-\frac{1}{2})k}}{\prod_{l \neq j}(1-q^{l-j})}.
\label{ferro}
\end{align}
\noindent
(v) Antiferromagnetic String Formation Probability:
\begin{align}
&P_{\rm a}(x,n) \nn \\
&\,
:=\bra E_x^{++} E_{x+1}^{--} 
\cdots  E_{x+n-1}^{\pm \pm} \ket
+\bra E_x^{--} E_{x+1}^{++} 
\cdots  E_{x+n-1}^{\mp \mp} \ket \nn \\
&
= \sum_{k=0}^{\infty}(-1)^k
\sum_{j=1}^n
\frac{q^{(x+j-\frac{1}{2})k}}{\prod_{l \neq j}(1-q^{l-j})} 
q^{\frac{n}{2}(x+\frac{n}{2})}\nn \\
&\,\,
\times 
\begin{cases}
   \(q^{\frac{x+\frac{n}{2}}{2}+\frac{(k+\frac{n+1}{2})^2}{2}}
      +q^{\frac{-x-\frac{n}{2}}{2}+\frac{(k+\frac{n-1}{2})^2}{2}}\)
   & \text{$n$: odd} \\
   \(q^{-\frac{n}{2}+\frac{(k+\frac{n}{2})^2}{2}}
   +q^{\frac{n}{2}+\frac{(k+\frac{n}{2})^2}{2}}
    \)
& \text{$n$: even}
\end{cases}.
\label{antiferro}
\end{align}
%

%
\begin{figure}[]
\includegraphics[width=\linewidth]{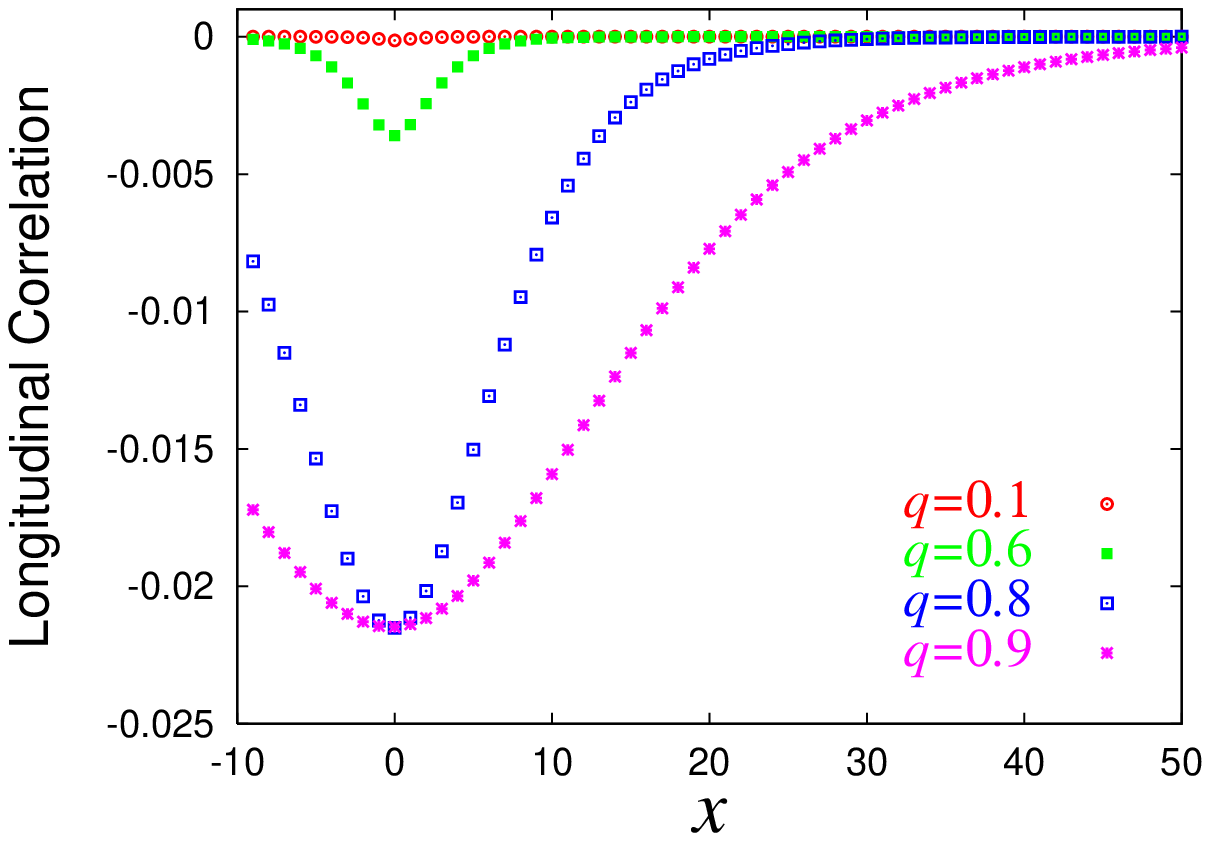}
\includegraphics[width=\linewidth]{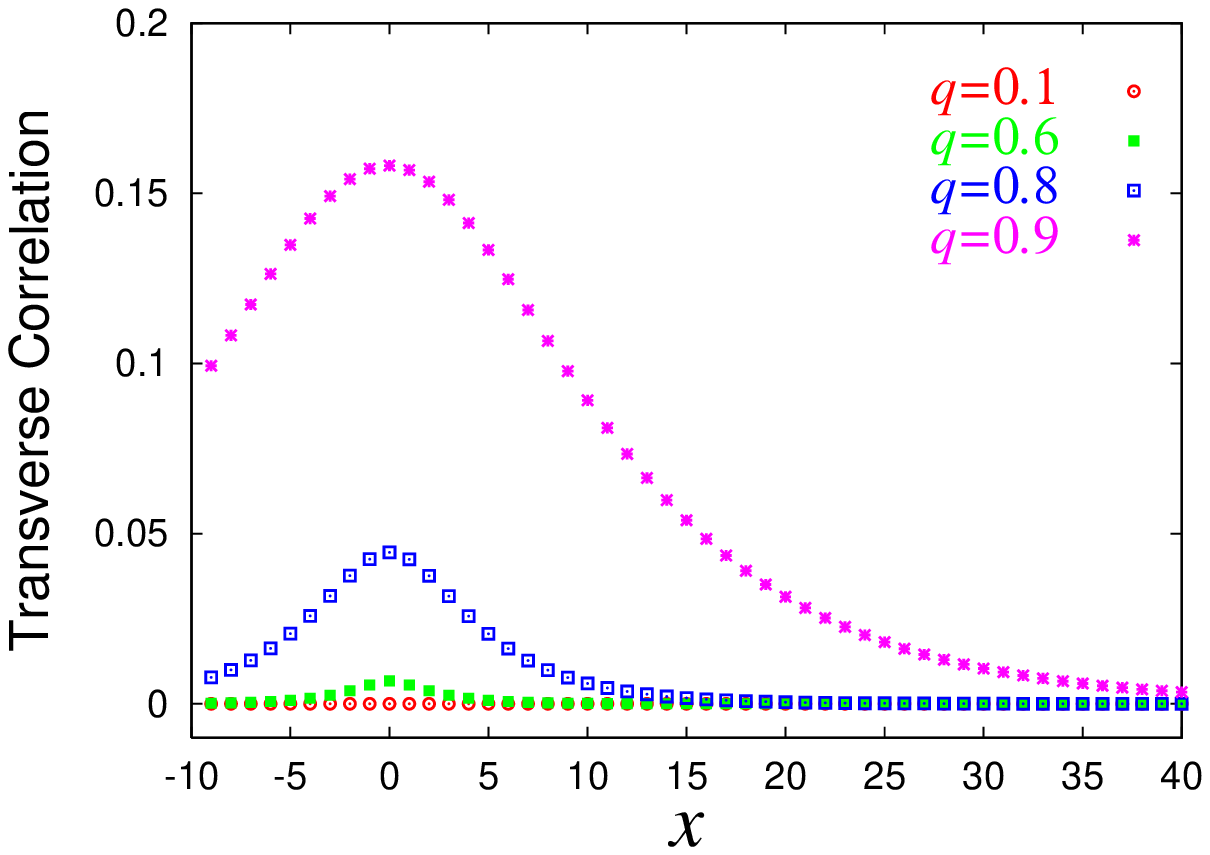}
\caption{Top: The longitudinal spin-spin 
correlation functions $C^{zz}(x_1,x):=
\bra \sigma_{x_1}^{z} \sigma_{x}^{z} \ket-
\bra \sigma_{x_1}^z \ket \bra \sigma_{x}^z \ket$ for $x_1=-10$ and $-10<x\le 50$.
Bottom: The transverse correlation functions $\bra \sigma_{x_1}^+ \sigma_{x}^- \ket$
for $x_1=-10$ and $-10<x\le 40$.}
\label{spin-spin}
\end{figure}

\begin{figure}
\includegraphics[width=\linewidth]{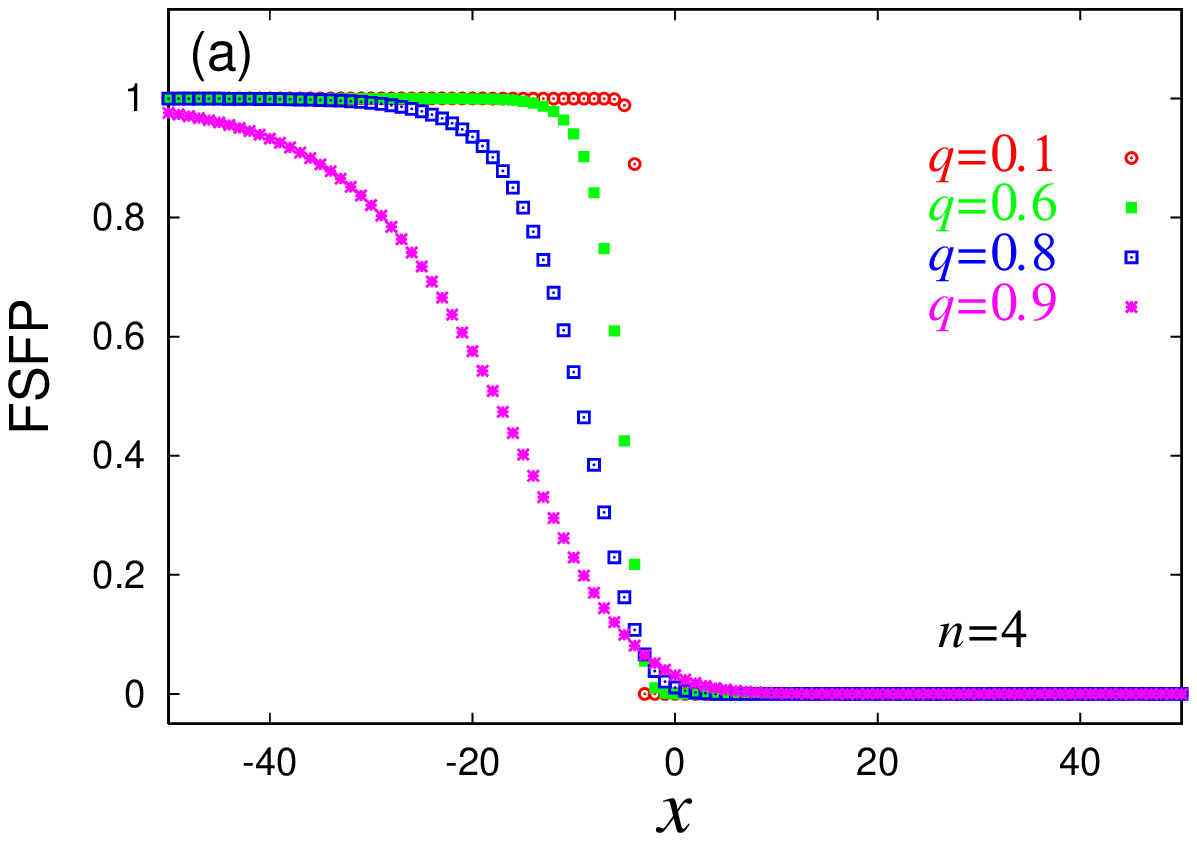}
\includegraphics[width=\linewidth]{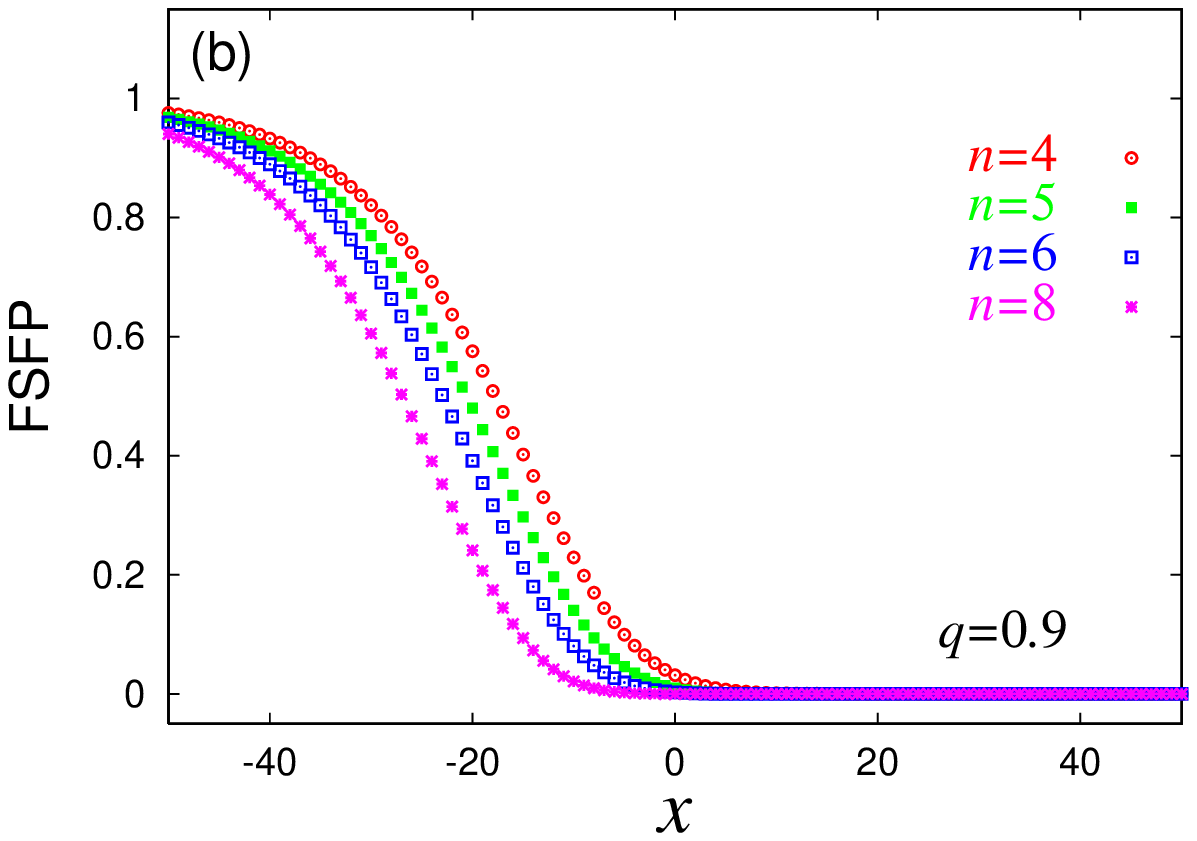}
\caption{The ferromagnetic string formation 
probability (FSFP)  $P_{\rm f}(x,n)$ for $n=4$ 
and various anisotropies $q$ (a), 
and for $q=0.9$ and various lengths $n$ (b).}
\label{FSFP}
\end{figure}

\begin{figure}[]
\includegraphics[width=\linewidth]{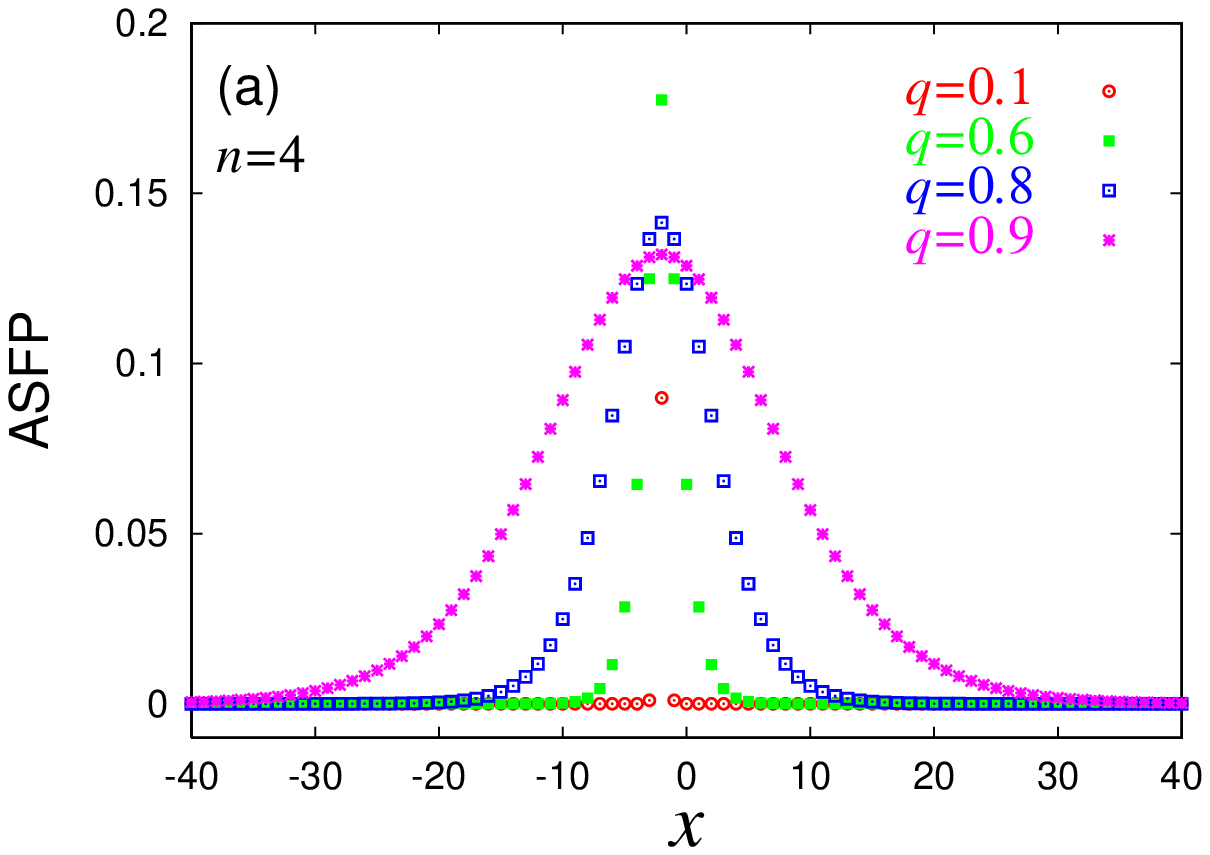}
\includegraphics[width=\linewidth]{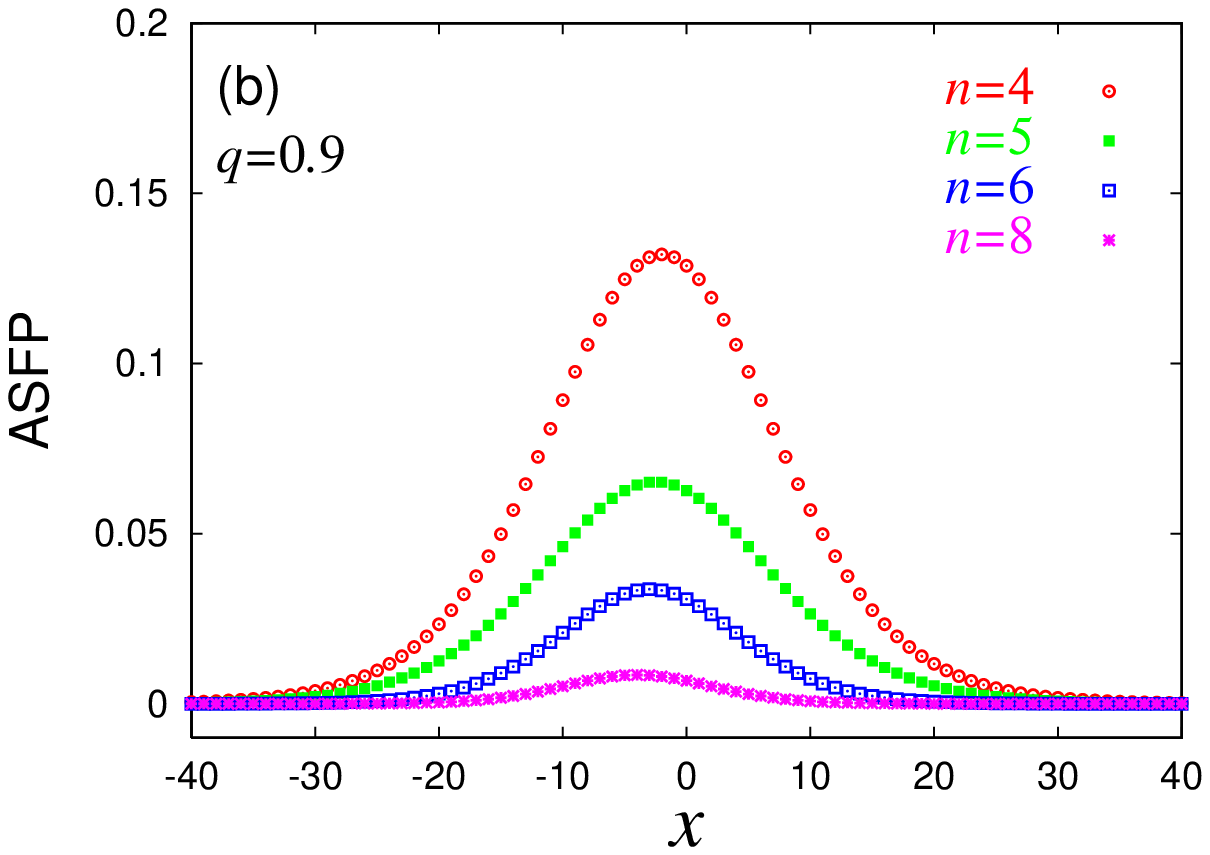}
\caption{The ferromagnetic string formation 
probability (ASFP)  $P_{\rm a}(x,n)$ for $n=4$ 
and various anisotropies $q$ (a), 
and for $q=0.9$ and various lengths $n$ (b).}
\label{ASFP}
\end{figure}

\begin{figure}[]
\includegraphics[width=\linewidth]{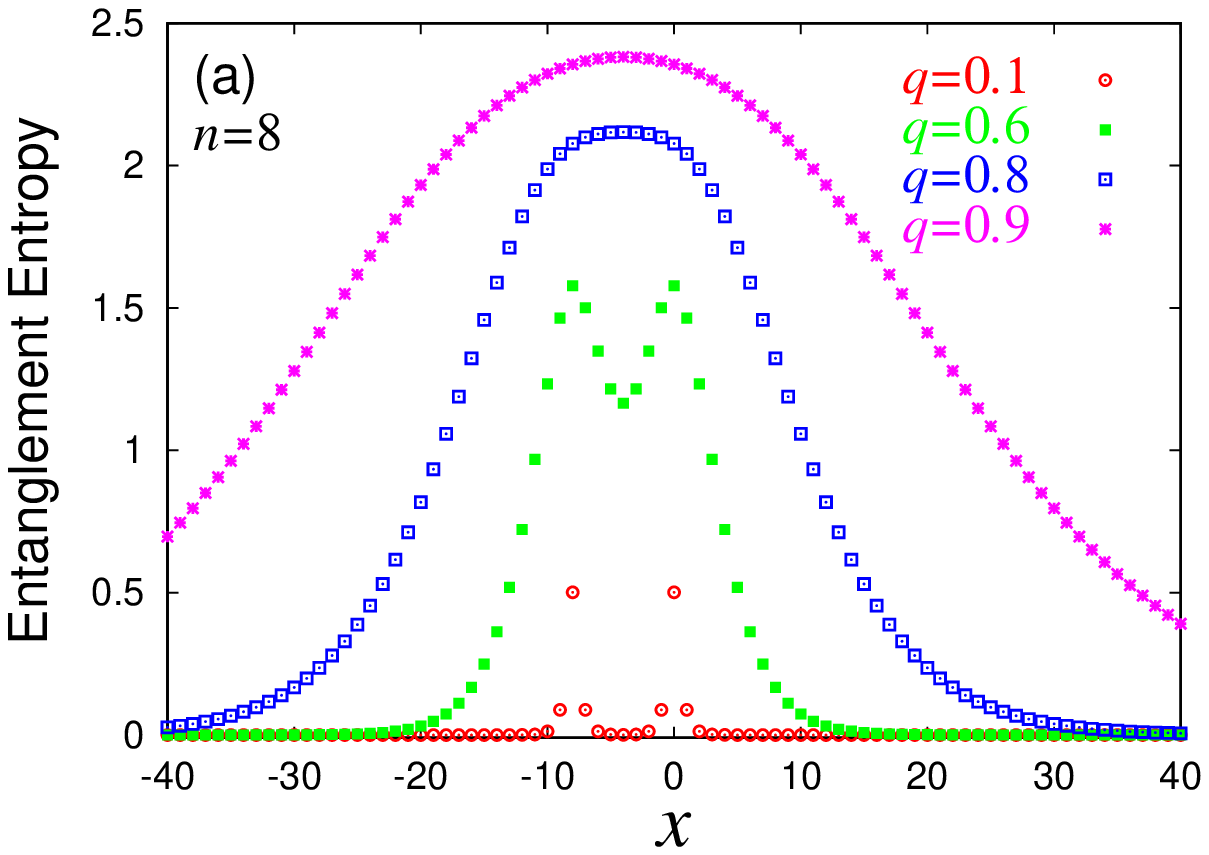}
\includegraphics[width=\linewidth]{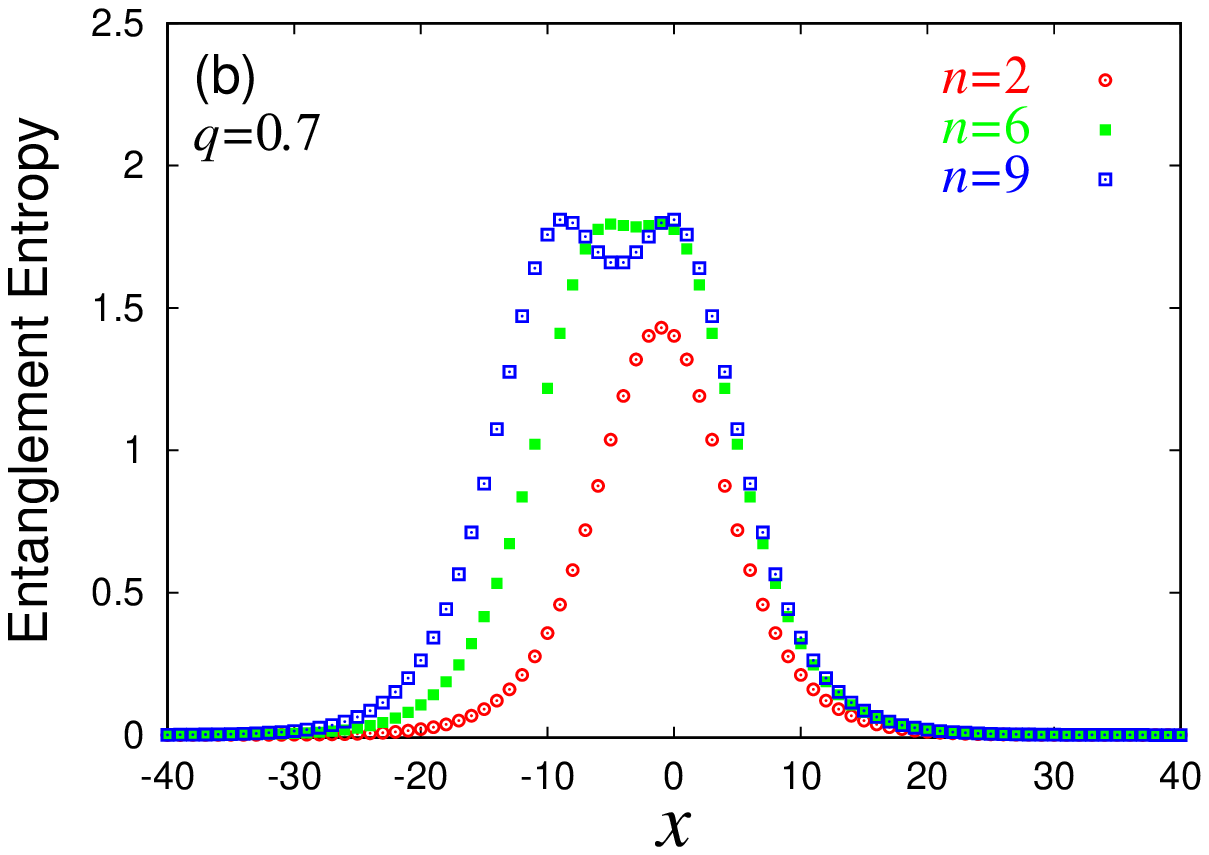}
\caption{The entanglement entropy  $S(x,n)$ for $n=8$ 
and various anisotropies $q$ (a), 
and for $q=0.7$ and various lengths $n$ (b).}
\label{entropy}
\end{figure}

FIG.~\ref{magnet} shows the magnetization 
$\bra \sigma_x^{z} \ket$. One sees that
the sign of the magnetization changes 
between $x=-1$ and $x=0$, which corresponds
to the fact that the center of the kink is located at
$x=-1/2$. One can also see that most of the 
spins are aligned up for $x\ll -1$ and aligned 
down for $x \gg 1$.
As decreasing of $q$ (or equivalently as 
increasing of the anisotropy parameter $\Delta$), the
slope of the magnetization curve around $x=0$ 
increases, and eventually will be infinite
at the Ising limit $q=0$ ($\Delta=\infty$).

The longitudinal and the transverse
spin-spin correlation functions
\begin{equation}
C^{zz}(x_1,x):=\bra \sigma_{x_1}^{z} \sigma_{x}^{z} \ket-
\bra \sigma_{x_1}^z \ket \bra \sigma_{x}^z \ket
\end{equation}
and $\bra \sigma_{x_1}^+ \sigma_{x}^- \ket$
are depicted in FIG.~\ref{spin-spin}
for the case $x_1=-10$ and various anisotropies.
Both the correlation functions have a peak around
$x=0$. This characteristic behavior reflects the
fact that  the sign of the magnetization 
changes around $x=0$ and almost all spins are aligned
at $|x|\gg 1$. These correlation functions decay
exponentially for $|x|\gg 1$ (see section \text{III} in detail).

The ferromagnetic (antiferromagnetic) 
string formation probability $P_{\rm f}(x,n)$ 
($P_{\rm a}(x,n)$)
is the probability that the spins
located in the region $[x,x+n-1]$ form a 
ferromagnetic (antiferromagnetic) string.
These correlation functions are depicted in 
FIG~\ref{FSFP} and FIG~\ref{ASFP}, respectively.
From FIG.~\ref{FSFP}-b and \ref{ASFP}-b, one can see 
$P_{\rm f}(x,m)>P_{\rm f}(x,n)$ and 
$P_{\rm a}(x,m)>P_{\rm a}(x,n)$ for $m<n$, as expected.
As shown in section IV, both the string formation probabilities
$P_{\rm f}(x,n)$ and $P_{\rm a}(x,n)$ 
exhibit Gaussian decay for large strings $n\gg 1$.
As the anisotropy parameter becomes larger,
the effect of Ising interaction becomes stronger 
than that of quantum fluctuation.
In the limit $q \to 0$ ($\Delta \to \infty$),
the spins for $x<0$ and $x>0$
are all aligned up and down, respectively
(cf. FIG.~\ref{magnet}).
This is reflected in
the slope in FIG. \ref{FSFP}-a becoming steeper,
and the peak in FIG. \ref{ASFP}-a becoming sharper,
as the anisotropy parameter becomes larger.
One can also calculate the entanglement entropy
$S(x, n)$, the von Neumann entropy of a 
subsystem $\[ x, x+1, \cdots , x+n-1  \]$.
It is defined as
\begin{eqnarray}
S(x, n)=-\textrm{tr} \rho(x, n) \log_2 \rho(x, n),
\end{eqnarray}
where $\rho(x, n)$ is the reduced density matrix
defined by tracing out the degrees of freedom
of  the environment outside the subsystem
$\[ x, x+1, \cdots , x+n-1  \]$:
\begin{align}
\rho(x, n)=&\textrm{tr}_{\rm E}
| \Psi_0 \ket \bra \Psi_0 |=
\big[ 
P_{\epsilon_1,..., \epsilon_n}^{\epsilon_1^{\prime},..., \epsilon_n^{\prime}}(x,n)
\big]_{\epsilon_j, \epsilon_j^{\prime}=\pm},
\end{align}
where
\begin{eqnarray}
P_{\epsilon_1,..., \epsilon_n}^
{\epsilon_1^{\prime},..., \epsilon_n^{\prime}}(x,n)
=
\bra 
\prod_{j=1}^n E_{x+j-1}^{\epsilon_j^{\prime} \epsilon_j}  
\ket.
\end{eqnarray}
Shown in FIG.~\ref{entropy} is the entanglement entropy. 
As $x \to \pm \infty$, the entanglement entropy 
of the kink ground state is asymptotically 0,
which is nothing but that of the 
ferromagnetic ground state \textit{up} and \textit{down}.
In FIG.~\ref{entropy}-a, we observe an intriguing phenomena that 
the peak of the entanglement entropy splits into two,
as decreasing the  parameter $q$ (or equivalently
as increasing the anisotropy parameter $\Delta$).
On the other hand, for fixed $q$, the same behavior
can also be observed in FIG. \ref{entropy}-b as
increasing the length of the subchain.
%
%
%
\section{asymptotics}
%
In this section, the asymptotic behaviors of the correlation 
functions derived in the preceding section are analyzed.

Let us first consider the spin-spin correlation functions. 
From \eqref{magnetization} and \eqref{z-correlation}, we find
\begin{align}
&\bra \sigma_{x_2}^z \ket \xrightarrow{x_2 \to +\infty} 
-1+2 q^{x_2+1}, \nn \\
&\bra \sigma_{x_1}^z \sigma_{x_2}^z \ket 
\xrightarrow{x_2 \to +\infty}  \nn \\
&
\,\,\,-\bra \sigma_{x_1}^z \ket
+(2+4 \sum_{k=0}^{\infty}
(-1)^{k+1}q^{\frac{k^2}{2}+k(x_1+\frac{3}{2})}
)q^{x_2+1}.
\end{align}
Thus we obtain
\begin{align}
&\bra \sigma_{x_1}^z \sigma_{x_2}^z \ket
-\bra \sigma_{x_1}^z \ket \bra \sigma_{x_2}^z \ket
\sim A^{zz}(x_1)q^{x_2+1}
\text{ for $x_2 \gg 1$}, \nn \\
&A^{zz}(x_1)=4 \sum_{k=0}^{\infty}
(-1)^k (1-q^k)q^{\frac{k^2}{2}+k(x_1+\frac{1}{2})}.
\end{align}
This shows that the longitudinal spin-spin correlation function
decays exponentially.
The asymptotics of the transverse spin-spin correlation 
function \eqref{pm-correlation} is also evaluated in 
the same manner:
\begin{align}
&\bra \sigma_{x_1}^+ \sigma_{x_2}^-  \ket
\sim A^{+-}(x_1)q^{\frac{1}{2}(x_2+\frac{1}{2})}
\text{ for $x_2 \gg 1$}, \nn \\
&A^{+-}(x_1)=\sum_{k=0}^{\infty}(-1)^k 
q^{\frac{(k+1)^2}{2}+(k+\frac{1}{2})(x_1+\frac{1}{2})},
\end{align}
which shows that the transverse spin-spin correlation function
also exhibits exponential decay.

Now, let us analyze the asymptotics of the string 
formation probabilities \eqref{ferro}.
Using the identity
\begin{align}
\sum_{j=1}^n \frac{1}{\prod_{l \neq j}^n(1-q^{l-j})}=1,
\end{align}
$P_{\rm f}(x,n)$ can be rewritten as
\begin{align}
&P_{\rm f}(x,n)=q^{nx+n^2}(1+B(n)), \nn \\
&B(n)=\sum_{k=1}^{\infty}(-1)^k
q^{\frac{k^2}{2}+(n+x-\frac{1}{2})k}
\sum_{j=1}^n \frac{q^{jk}}{\prod_{l \neq j}^n(1-q^{l-j})}.
\end{align}
Since 
\begin{align}
|B(n)| &< 
\sum_{k=1}^{\infty}
q^{\frac{k^2}{2}+(n+x-\frac{1}{2})k}
\sum_{j=1}^n \frac{q^{jk}}{\prod_{l \neq j}(1-q^{l-j})} \nn \\
&<
\sum_{k=1}^{\infty}
q^{(n+x-\frac{1}{2})k}
\sum_{j=1}^n \frac{1}{\prod_{l \neq j}(1-q^{l-j})} \nn \\
&=\sum_{k=1}^{\infty}
q^{(n+x-\frac{1}{2})k},
\end{align}
then 
\begin{align}
P_{\rm f}(x,n) \sim q^{nx+n^2} \text{ for $ n \gg 1$}.
\end{align}
This means that the ferromagnetic string formation
probability shows Gaussian decay for large strings.
Note here that  similar Gaussian behaviors are 
also seen in the {\it antiferromagnetic} ground state
\cite{KMST,KMST2,KLNS}.

Finally we explicitly write down the asymptotics of 
the antiferromagnetic string formation probability
\eqref{antiferro}:
\begin{align}
P_{\rm a}(x,n)  & \sim
q^{\frac{nx}{2}+\frac{3n^2}{8}} \nn \\
&\times
\begin{cases}
\(q^{\frac{x}{2}+\frac{4n+1}{8}}
  +
    q^{-\frac{x}{2}+\frac{-4n+1}{8}}\)
& \text{$n$: odd} \\
   \(q^{-\frac{n}{8}}
     +
    q^{\frac{n}{8}}
\)
& \text{$n$: even}
\end{cases},
\end{align}
for $n \gg 1$. 
\section{conclusion}
In this paper, the density matrix in the kink 
ground state of the ferromagnetic spin-1/2 XXZ 
chain has been exactly calculated.
From this expression, the longitudinal
and transverse spin-spin  correlation functions, 
and the ferromagnetic and the antiferromagnetic
string formation probability for arbitrary distances
and arbitrary interaction strengths have been systematically
calculated.
Analyzing them, we find that the spin-spin correlation
functions decay exponentially for large distances,
while the string formation probabilities show
Gaussian decay for large strings.
We have also calculated the entanglement entropy
and observed the change of shape with 
the increase of the anisotropy parameter
or the length of the subchain.
\begin{acknowledgments}
This work  was partially  supported by Global COE Program
(Global Center of Excellence for Physical Sciences Frontier)
and Scientific Research (B) No. 18340112
from MEXT, Japan.
\end{acknowledgments}

\appendix
\section*{Appendix: Proof of $\eqref{exp2}$}

Let us show \eqref{exp2}. Using \eqref{GF},
one obtains
\begin{align}
&\bra \Psi(z) | \prod_{j=1}^n E_{x_j}^{\epsilon_j^{\prime} \epsilon_j} | \Psi(z) \ket \nn \db
&=
\prod_{j=1}^n \frac{1}{1+w \zeta_j}
\prod_{j=1}^n (w \zeta_j)^{\[ \epsilon_j^{\prime} \epsilon_j \]}
(-wq^{\frac{1}{2}};q)_{\infty}
(-w^{-1}q^{\frac{1}{2}};q)_{\infty}
 \nn \db
&= 
g_w^n(\zeta_1, ..., \zeta_n)
\prod_{j=1}^n (w \zeta_j)^{\[ \epsilon_j^{\prime} \epsilon_j \]}
\frac{1}{(q;q)_{\infty}}
\sum_{j=-\infty}^{\infty}w^j q^{\frac{j^2}{2}}, \label{app} \nn \db
&g_w^n(\zeta_1, ..., \zeta_n)
:=\prod_{j=1}^n \frac{1}{1+w \zeta_j}. 
\end{align}
Expressing the Laurent expansion of $g_w^n(\zeta_1, ..., \zeta_n)$ as
\begin{equation}
g_w^n(\zeta_1, ..., \zeta_n)=\sum_{j=0}^{\infty}w^j \lambda_j^{(n)}
(\zeta_1,..., \zeta_n), \label{laurent}
\end{equation}
we find 
$\lambda_j^{(n)}(\zeta_1,..., \zeta_n)$  satisfies
the following recursion relation,
\begin{align}
\lambda_k^{(n+1)}(\zeta_1,..., \zeta_{n+1})
=\sum_{j=0}^k(-1)^{k-j}\zeta_{n+1}^{k-j}
\lambda_j^{(n)}(\zeta_1,...,\zeta_n).
\label{recursive}
\end{align}
To prove \eqref{exp2} is to show that $\lambda_j^{(n)}(\zeta_1,..., \zeta_n)$ is
\begin{equation}
\lambda_j^{(n)}(\zeta_1,..., \zeta_n)=(-1)^j \sum_{l=1}^n 
\frac{\zeta_l^{j+n-1}}{\prod_{i \neq l}(\zeta_l-\zeta_i)}. \label{lambda}
\end{equation}
Let us show this by induction.
It is obvious that \eqref{lambda} holds for $n=1$.
Suppose it holds for $n$.
Then from 
\eqref{recursive}, $\lambda_k^{(n+1)}
(\zeta_1,...,\zeta_{n+1})$
can be calculated as follows.
\begin{align}
&\lambda_k^{(n+1)}(\zeta_1,...,\zeta_{n+1}) \nn \\
&\,\,=(-1)^k \zeta_{n+1}^k \sum_{j=0}^k \sum_{l=1}^n \zeta_l^{n-1}
\left( \frac{\zeta_l}{\zeta_{n+1}} \right)^j 
\frac{1}{\prod_{\substack{i=1 \\  i \neq l}}^n (\zeta_l-\zeta_i)
} \nn \\
&\,\,=(-1)^{k} \sum_{l=1}^n \frac{\zeta_l^{n+k}-\zeta_l^{n-1} \zeta_{n+1}^{k+1}}
{\prod_{\substack{i=1 \\  i \neq l}}^{n+1} (\zeta_l-\zeta_i)} \nn \\
&\,\,=(-1)^{k} \sum_{l=1}^{n+1} \frac{\zeta_l^{n+k}}
{\prod_{\substack{i=1 \\  i \neq l}}^{n+1} (\zeta_l-\zeta_i)}. \label{ind}
\end{align}
In the last equality, we used 
\begin{eqnarray}
\sum_{j=1}^n \frac{\zeta_j^{n-2}}{\prod_{k \neq j}(\zeta_j-\zeta_k)}=0
\quad \text{for $n \ge 2$}.
\label{ass}
\end{eqnarray}
From \eqref{ind},
we can see \eqref{lambda} holds for $n+1$,
which means \eqref{lambda} holds for any $n$.
Thus, from \eqref{app}, \eqref{laurent} and \eqref{lambda},
we obtain \eqref{exp2}.
%
%

\end{document}